\documentclass[prl,twocolumn,floatfix,showpacs]{revtex4}
\usepackage{graphics}
\usepackage{dcolumn}
\usepackage{bm}
\usepackage{epsfig}
\usepackage{pictex}

\begin{document}
\draft
\preprint{HEP/123-qed}
\title{Filling dependence of a new type of charge ordered liquid on a triangular lattice system}
\author{Chisa Hotta
\footnote{\vspace*{-10mm} electric address: chisa@phys.aoyama.ac.jp} 
and Nobuo Furukawa}
\address{Aoyama-Gakuin University, 5-10-1, Fuchinobe, Sagamihara, Kanagawa 229-8558, Japan}
\date{June 31, 2006 (submitted to  J. Phys.: Cond. Matt.)}
\begin{abstract}
We study the recently reported new characteristic gapless charge ordered state 
in a spinless fermion system on a triangular lattice 
under strong inter-site Coulomb interactions. 
In this state 
the charges are spontaneously divided into solid and liquid component, 
and the former solid part aligns in a Wigner crystal manner while 
the latter moves among them like a pinball. 
We show that such charge ordered liquid is stable 
over a wide range of filling, $1/3<n<2/3$, 
and examine its filling dependent nature. 
\end{abstract}
\maketitle
\section{Introduction}
Ever since the concept of frustration-induced ``spin liquids'' is 
introduced to the triangular lattice systems\cite{anderson}, 
the theoretical search for a special state without any types of 
long range ordering has been proceeded in the the strongly correlated systems 
under geometrical frustration\cite{sorella,imada}. 
Such liquids are now being explored experimentally in the materials 
like NiGa$_2$S$_4$\cite{nakatsuji} and 
$\kappa$-ET$_2$Cu$_2$(CN)$_3$\cite{kappa}. 
However, the competition among several orders towards 
their destruction often makes the system 
reconcile with another different type of ordering. 
Simplest examples are the spins on frustrated lattices forming 
dimers\cite{miyahara}, plaquettes\cite{plaquette}, etc. 
Most recent ones are reported on hard core bosons 
of a nematic ordering in frustrated square lattice\cite{nematic} 
as well as of supersolids in the triangular lattice\cite{supersolid}. 
We also proposed a frustration induced novel charge liquid 
in a triangular lattice denoted as a "pinball liquid"\cite{pinball}, 
which turned out to have very similar character with the above supersolids 
despite the difference in the statistics. 
This state has a coexistence of a Wigner crystal-type of solid and 
itinerate liquid type of charges so that a phase separation is 
expected without any domain structures. 
In this paper we examine the filling dependent nature of this liquid in detail 
in the spinless fermion system and confirm the strong coupling picture. 
\section{Model}
The present spinless fermion system is described by the $t$-$V$ model hamiltonian, 
\begin{eqnarray}
{\cal H} &=&\sum_{\langle i,j \rangle} \Big(
         -t c^\dagger_i c_j + {\rm h.c.} + V n_i n_j\Big), 
\label{tvham}
\end{eqnarray}
where $c_j$ denotes the annihilation operator of fermions and 
$n_j(=c^\dagger_j c_j)$ is its number operator. 
The summation is given over the nearest neighbor(nn) pair sites 
on a triangular lattice with indices $\langle ij \rangle$. 
Together with the strong coupling argument, 
we perform the exact diagonalization on $N=4\times 6$ cluster 
with periodic boundary conditions, which could fully describe 
both the three fold and two fold type of periodicity 
which reside in the present system. 
Since the size dependence of the energy is confirmed to be very small 
and also the correlation function shows a well converged character 
within a relatively short range 
in comparison with the parallel works\cite{pinball,supersolid}, 
the present results are enough qualified to 
support the physical picture we present. 
\begin{figure}[t]
\begin{center}
\includegraphics[width=8cm]{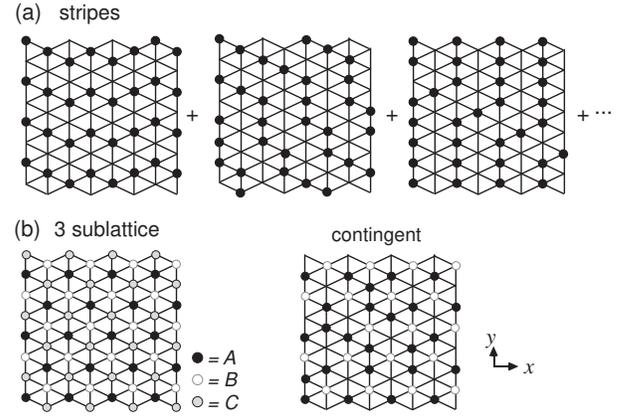}
\end{center}
\caption{
(a) Schematic illustration of the representative 
classical ground states, (a) two-fold stripes, 
(b) basic three sublattice structure and one of its contingent.
}
\label{f1}
\end{figure}
\begin{figure}[t]
\begin{center}
\includegraphics[width=8.5cm]{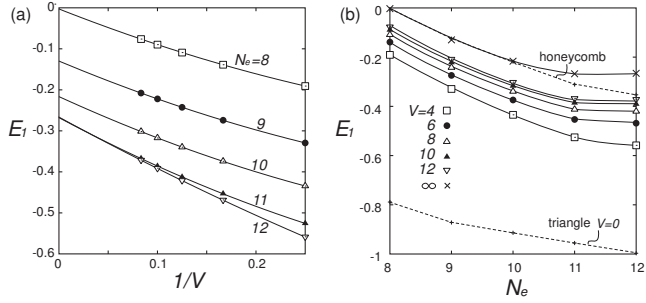}
\end{center}
\caption{Kinetic energy gain, $E_1=(E_Q-E_C)/N$, 
in the formation of a pinball liquid at $N=24$ with $t=1$
as a function of (a) $1/V$
and (b)$N_e$, where $N_e$=8,12 corresonds to $n$=1/3, 1/2, respectively. 
The fitting by the power of $1/V$ is given in (a), whose 
results are reflected in (b) as those of $V=\infty$. 
The band energy gain of the free electrons on a 
corresponding honeycomb($N=16$) and triangular ($N=24$) lattices 
are shown together for comparison. 
}
\label{f2}
\end{figure}
\begin{figure}[t]
\begin{center}
\includegraphics[width=8cm]{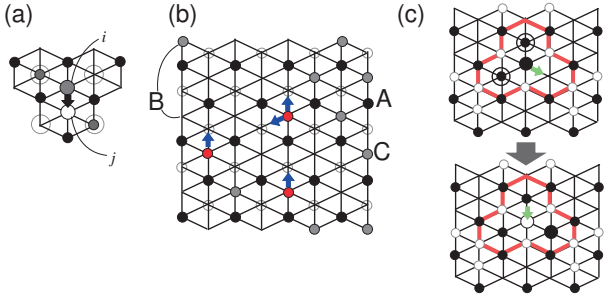}
\end{center}
\caption{
(a) Example of two basis that is allowed to mix by $t$. 
The ball can move from site $i$ to $j$ when the extra two neighbors of site $i$
with big circles together have the same number of balls with those 
of the site $j$. 
(b) Representative basis of the three sublattice states at $n=0.5$. 
Filled black circle and gray or red circles denote the "pin" and "ball", respectively, 
and the open circle the empty $C$-sites. 
The "balls" on $C$-site can move to neighboring $B$-site and vise versa 
 when condition (a) is fulfilled. Arrows show the possible movement. 
(c) Example of the case when the "pin"(big black circle) on A-site move to its neighbor. 
When the black circle moves away from the configuration 
shown in the bottom, the two "balls" marked with big circle can no longer move. 
If one of these "balls" refill the original "pin", then other balls can move away. 
}
\label{f3}
\end{figure}
\begin{figure}[t]
\begin{center}
\includegraphics[width=8.5cm]{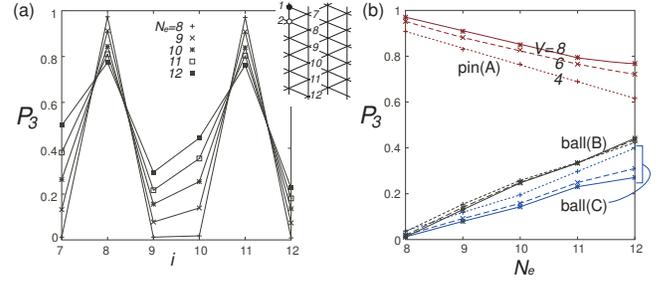}
\end{center}
\caption{
(a) three-body correlation function, $P_3(i)=\langle n_1 (1-n_2) n_i \rangle$, 
along the second chain at several fillings between $1/3\geq n \neq 1/2$. 
$N_e=8,12$ correspond to 1/3 and 1/2-fillings, respectively. 
(b) Amplitude of $P_3$ as a function of $N_e$ 
at sites $i=4,5,6$ which belong to the $A$, $B$, $C$-sublattices, respectively. 
The difference between pins and balls become small near half-filling. 
}
\label{f4}
\end{figure}
\section{Strong coupling theory}
In the strong coupling limit, $t=0$, the system is 
identical with the classical Ising $S$=1/2 spin system on a triangular lattice. 
The ground state is disordered due to macroscopic number of degenerate states\cite{wannier} 
which are classified into two groups according to their periodicity 
as shown in Fig.~\ref{f1}; 
one portion is composed of two-fold striped states 
which have staggered alignment of charges in any one of three directions 
but is disordered in other two directions. This state has an order $\sqrt{N}$-degeneracy. 
The others have a three sublattice structure represented by Fig.~\ref{f1}(b). 
Here, as long as $A$ and $C$ sublattices are filled and empty, respectively, 
the configuration of the $B$-sites is chosen arbitrary. 
In addition, there are large number of contingents which include 
the irregular structure on $A$ and $B$ sublattices as shown in Fig.~\ref{f1}(c) 
but still has a same binding energy. 
Including all of them, 
the three sublattices can afford any fillings over $1/3<n<2/3$ 
without changing the binding energy, in sharp contrast to 
the strictly half-filled stripes. 
\par
The introduction of $t \neq 0$ drives the classical disordered system 
to a new ordered phase which we call a pinball liquid\cite{pinball}. 
For simplicity, we start from the $n=1/3$ case 
which has a well known unique ground state 
where only the $A$-sublattice is filled. 
Then the extra binding energy $2V$ 
is required to move a single charge from one of the $A$-sites, 
thus the "pins" on $A$-site form a solid. 
Next we add a single extra charge on one of the empty sites. 
This charge can move to another nearest empty site without 
losing the binding energy. 
Therefore once the filling is off $n=1/3$, the system discontinuously 
becomes gapless. 
The simplest picture of the pinball liquid is that 
the extra charge moves around like a "ball" avoiding the "pins" on $A$ sublattice. 
The kinetic energy gain per single "ball" 
corresponds to the bandwidth of the honeycomb lattice which is obtained after 
depleting the $A$ sublattice(pinned sites) from the original lattice. 
With increasing filling the energy gain will be affected by 
the correlation among "balls". 
We plot in Fig.~\ref{f2} the estimation of 
$E_1=(E_Q-E_C)/N$, where $E_Q$ is numerically obtained 
by the diagonalization of eq.(\ref{tvham}) 
and the classical binding energy is given by $E_C=3V N_e$, 
where $N_e$ denotes the electron number. 
The kinetic energy gain, $|E_1|$, increases with increasing number of filling, 
i.e. the number of "balls" that can move. 
In Fig.~\ref{f2}(a) we fit 
the data by the power of $1/V$ as, $E_1=c_0+c_1/V+c_2/V^2$, 
where $c_j$ are fitting constants and $c_0$ gives the kinetic energy gain 
in the strong coupling limit. 
The obtained results at $V=\infty$ are given as a function of filling in Fig.~\ref{f2}(b) 
together with $E_1$'s at finite $V$. 
In the strong coupling limit, $E_1(V=\infty)$ seems to follow 
the band energy of the honeycomb lattice at $N_e<10$ which supports 
the simplest pinball liquid picture we discussed above in the dilute case. 
As the filling approaches $n=1/2$, $E_1$ deviates from the honeycome one and saturates. 
This is because the correlation among "balls" increases to avoid the loss of $V$ 
as they becomes dense. Then the balls can no longer move independently 
with each other and the wave functions become relatively localized 
so that the kinetic energy gain do not increase much with $N_e$. 
\par
Although we see shortly that the pinball liquid is dominant over $1/3< n \leq 1/2$, 
its stability is still the most difficult to be guaranteed at half-filling 
because the largest number of contingents exists 
that might destroy the "pins" to some extent. 
We remind that the charges on $B$ and $C$ sites 
can hop by $t$ to its neighboring site 
as in Fig.~\ref{f3}(a) without the loss of binding energy when 
the neighboring population number of that charge remains unchanged. 
The example is given in Fig.~\ref{f3}(b) where 
some of the balls can move in the direction indicated by arrows. 
This rule also holds for the charges on $A$ site 
in the irregular contingent states, 
e.g. when three neighboring hexagons have special kind of configuration 
as shown in Fig.~\ref{f3}(c). 
Fortunately, in this case the contingent cannot destroy the "pins" 
completely. 
This is because the charges which was originally a "pin" cannot go more 
than one site away without freezing the configuration of 
original two neighbors. 
Otherwise, one of these neighbors refills the original "pin" and 
the "balls" can move again smoothly. 
Therefore although the local fluctuation squeeze the amplitude of "pins" 
they remain relatively stable\cite{pinball} and the collective destruction 
of "pins" does not take place. 
Considering the particle-hole symmetry, the same discussion holds 
by regarding the $C$-sublattice(hole) as "pins". 
Therefore, the "pins" and "balls" are not completely fixed and free, 
respectively, at half-filling. 
\section{Correlation functions}
To further compare the stability of the pinball liquid under different fillings, 
we present the numerical estimation of order parameters for 
several number of charges $N_e$=8-12 at $N=24$, 
which correspond to the fillings, $n$=1/3-1/2. 
As discussed elsewhere, the pinball liquid is characterized by 
a three-body correlation\cite{pinball}. 
Figure~\ref{f4} shows such function, $P_3(i)=\langle n_1 (1-n_2) n_i \rangle$, at $t=1$. 
As we see in Fig~\ref{f4}(a), the amplitude of "pins"($A$) 
near $n$=1/3 is almost 1 which decays only little with distance. 
This sturdily supports the strong coupling picture and is also consistent 
with the fact that the kinetic energy gain in Fig.~\ref{f2}(a) 
at low fillings is almost identical with the honeycomb lattice band energy. 
With increasing $N_e$ the oscillation 
amplitude of $P_3$ becomes suppressed and has larger decay. 
It is more clearly displayed in Fig~\ref{f4}(b) as a function of $N_e$ 
for several choices of $V$. 
However, the amplitude of "pins" are always larger for larger $V$, 
and we expect a long range order in the bulk limit at least in the 
very strong coupling region even at half-filling. 
\section{Summary}
We displayed the filling dependence of the 
new gapless charge liquid denoted as pinball liquid, 
which was reported very recently at $1/3 < n \leq 1/2$. 
Due to the particle-hole symmetry, the same results are obtained 
for $1/2 \leq n < 2/3$ by just regarding holes as "pins". 
Despite the anticipation that the increasing number of particles 
towards half-filling might drive the "pins" towards destruction, 
the kinetic energy gain and the correlation function indicates that 
the pinball picture still holds while 
the balls become strongly correlated. 
\section{Acknowledgments}
We are grateful to K. Kubo, S. Miyahara, and C. Yasuda for discussions.

\end{document}